\documentclass[pra,aps,twocolumn,longbibliography,preprintnumbers,superscriptaddress,nofootnotebib,10pt]{revtex4-2}

\usepackage[dvipsnames]{xcolor}
\definecolor{mediumpersianblue}{rgb}{0.0, 0.4, 0.65}
\definecolor{persianred}{rgb}{0.8, 0.2, 0.2}
\definecolor{darklavender}{rgb}{0.45, 0.31, 0.59}
\usepackage[normalem]{ulem}
\usepackage[colorlinks=true,citecolor=mediumpersianblue,linkcolor=mediumpersianblue, urlcolor=mediumpersianblue]{hyperref}

\usepackage{graphicx} 
\usepackage{enumitem}
\usepackage{braket}
\usepackage{amsfonts}
\usepackage{fontawesome}
\usepackage{amsmath}
\usepackage{dsfont}
\usepackage{listings}

\newcommand{\numqubits}{L}  
\newcommand{\qubitidx}{l}   
\newcommand{\maxintlen}{{\ell}} 
\newcommand{\numgates}{N}   
\newcommand{\numcliff}{{N_{\rm C}}} 
\newcommand{\nummagic}{{N_{\rm nC}}} 
\newcommand{\numlayers}{D}  
\newcommand{\mrr}{\beta}    
\newcommand{\numrotdecomp}{\nummagic}

\usepackage{listings}
\usepackage{xcolor}
\definecolor{lightcodegray}{rgb}{0.97, 0.97, 0.97}
\definecolor{softborder}{gray}{0.8}

\lstdefinestyle{pythonstyle}{
    language=Python,
    backgroundcolor=\color{lightcodegray},
    commentstyle=\color{ForestGreen}\itshape,
    keywordstyle=\color{NavyBlue}\bfseries,
    numberstyle=\tiny\color{gray},
    stringstyle=\color{BrickRed},
    basicstyle=\ttfamily\small,
    breakatwhitespace=false,
    breaklines=true,
    captionpos=b,
    keepspaces=true,
    numbers=left,
    numbersep=5pt,
    showspaces=false,
    showstringspaces=false,
    showtabs=false,
    tabsize=2,
    frame=single,
    framerule=0.3pt,
    rulecolor=\color{softborder}
}

\newcommand{\mpstab}{\texttt{mpstab}}

\begin{document}

\title{\texttt{MPStab}: an hybrid stabilizers tensor-network quantum circuit simulator}

\author{Giulio Crognaletti}
\affiliation{Department of Physics, University of Trieste, Strada Costiera 11, 34151 Trieste, Italy}
\affiliation{Istituto Nazionale di Fisica Nucleare, Trieste Section, Via Valerio 2, 34127 Trieste, Italy}
\affiliation{Department of Physics, University of Helsinki, P.O. Box 43, FI-00014 Helsinki, Finland}
\author{Mattia Robbiano}
\affiliation{Department of Physics, University of Helsinki, P.O. Box 43, FI-00014 Helsinki, Finland}
\author{Michele Grossi}
\affiliation{European Organization for Nuclear Research (CERN), Geneva, Switzerland}
\author{Matteo Robbiati}
\affiliation{Department of Microtechnology and Nanoscience, Chalmers University of Technology, SE-412 96 Gothenburg, Sweden}

\begin{abstract}
The development of techniques for simulating quantum systems using classical computers is 
a paramount task for two primary reasons: \textit{i)} there exist configurations for which 
classical computers are remarkably effective and will continue to be so, and \textit{ii)} 
exploring the limits of classical computation facilitates the identification of the 
regimes of competence for quantum computers. In this work, we present \texttt{mpstab}, 
a quantum circuit simulator based on a hybrid formalism combining stabilizers and 
tensor networks, recently introduced in \cite{Mello_2024}. We present the package, its 
core functionalities, and explore its performances in a few interesting simulation regimes.
\end{abstract}

\maketitle

\tableofcontents

\section{Introduction}

Simulating quantum systems on classical computers is a fundamental challenge in quantum information science. 
The most straightforward approach, statevector simulation, represents the quantum state as a vector and applies 
gates sequentially in the Schrödinger picture. This approach underlies several mature frameworks, from 
general-purpose platforms such as Qiskit~\cite{qiskit}, Cirq~\cite{cirq}, PennyLane~\cite{pennylane}, and Qibo~\cite{Qibo}, 
to performance-oriented engines such as Qulacs~\cite{qulacs}.
This method is simple and exact, making it ideal for validating 
small quantum algorithms and benchmarking. However, the state space dimension grows exponentially with system size, 
rendering this approach intractable beyond a few dozen qubits. Even with modern improvements such as just-in-time 
compilation~\cite{qibojit} or distributed computing~\cite{tejedor2025distributedquantumcircuitcutting}, this exponential scaling remains an insurmountable bottleneck for simulating larger systems.

\begin{figure}[t!]
    \centering
    \includegraphics[width=0.9\linewidth]{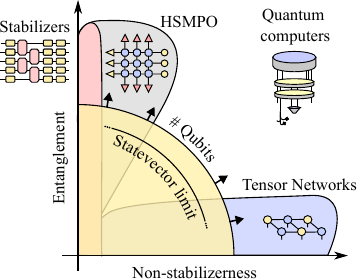} 
    \caption{\label{fig:classical_simulators} A qualitative representation of classical simulator realms. Laying on the entanglement axis, stabilizers simulators allow representing pure-Clifford circuits at large scale. On the magic axis, tensor networks consent simulation of large systems if entanglement growth remains under control. Mixed approaches, like the hybrid stabilizer-matrix product operator (HSMPO) implemented here, push the boundaries of the classical simulators closer to the bijector of the plane, which corresponds to the quantum computers realm.}
\end{figure}

Therefore, researchers have developed structured algorithms that exploit specific properties of quantum circuits to extend the classical simulation boundary. These approaches succeed by restricting attention to special dynamical regimes where the full state space dimension is not needed. In practice, as sketched in Fig.~\ref{fig:classical_simulators}, the difficulty of simulating many-body quantum dynamics stems from two competing sources of computational complexity: \textit{entanglement} and \textit{non-stabilizerness}. Quantum computers become necessary when both resources are extensively present, and classical simulators fail. Indeed, on one hand, tensor networks and particularly matrix product states (MPS), excel at simulating systems with local interactions. They are especially powerful for short-time dynamics and ground state calculations of local Hamiltonians where entanglement remains bounded. However, in generic quantum evolution, rapid entanglement growth forces an exponential increase in the tensor-network bond dimension, leading to a breakdown of efficiency~\cite{Orus_2014}. This picture has matured over the years into a rich software ecosystem including Quimb~\cite{quimb}, ITensor.jl~\cite{itensor}, TenPy~\cite{tenpy} and Quantum Tea~\cite{qmatchatea}.

On the other hand, stabilizer methods sit at the opposite extreme. They bypass the entanglement barrier, enabling exact polynomial-time simulation of highly entangling circuits~\cite{Aaronson_2004} by tracking Pauli stabilizers with efficient tableau arithmetic, an approach whose reference implementation is Stim~\cite{stim}. Yet, they lack universality, supporting only \emph{Clifford} gates that map Pauli strings into other Pauli strings. Because non-Clifford gates map Pauli operators into complex superpositions, extending these stabilizer simulations to accommodate operations like T gates or arbitrary angle rotations generally incurs in an exponential cost.

Several hybrid and specialized approaches have emerged to address broader dynamical regimes. \textit{Pauli propagation} methods evolve observables (rather than the full state) and maintain sparsity through truncation, proving effective for noisy circuits and systems with arbitrary connectivity~\cite{Rudolph_2025}. \textit{Lie-algebraic simulations}~\cite{Goh_2025} exploit circuits' underlying group structure for efficient evolution when the dynamical Lie algebra remains small. \textit{Neural Quantum States}~\cite{Lange_2024} leverage neural networks to parameterize volume-law entangled states inaccessible to tensor networks. A closely related line of work generalizes the tableau formalism itself to interface directly with tensor networks, enabling universal circuit simulation while retaining the efficiency of stabilizer tracking on the Clifford part~\cite{Masot_Llima_2024}.
Among these hybrid approaches, the hybrid stabilizer-matrix product operator (HSMPO) framework~\cite{Mello_2024} was recently introduced to bridge tensor networks and stabilizer methods. Exploiting the Pauli-preserving property of Clifford conjugations we can push Clifford gates past non-Clifford rotations and represent the result as compact tensor networks with fixed bond dimension. This method achieves polynomial-time simulation for circuits dominated by Clifford gates yet interspersed with moderate non-Clifford rotations, that neither tensor networks nor stabilizers can efficiently handle alone.

It is in this context that we introduce \texttt{mpstab}, a Python library implementing the HSMPO formalism for hybrid quantum circuit simulation.

In the following, we first summarize the HSMPO formalism in Sec.~\ref{sec:hsmpo}. Then, we introduce \texttt{mpstab} in Sec.~\ref{sec:mpstab}, 
describing its main utilities and showcasing usage examples alongside performance benchmarks against other state-of-the-art simulators. \\

\faGithub\,\, Open-source package: \\ \href{https://github.com/MatteoRobbiati/mpstab}{https://github.com/MatteoRobbiati/mpstab.} \\

\faBook \,\, Documentation: \\ \href{ https://matteorobbiati.github.io/mpstab/}{ https://matteorobbiati.github.io/mpstab.}

\section{Hybrid stabilizer-MPO}
\label{sec:hsmpo}

Here we summarize the workings of HSMPO simulations, a technique introduced in Ref.~\cite{Mello_2024} which attempts to bridge tensor networks and stabilizer methods, highlighting the role of both components into the algorithm.

\subsection{Circuit decomposition}
\begin{figure}[ht]
    \centering
    \includegraphics[width=1\linewidth]{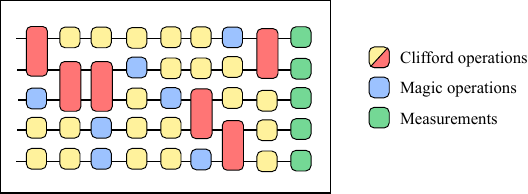} 
    \caption{\label{fig:stab_ent}Any quantum circuit can be written in terms of 
    Clifford operations (red and yellow gates) alternated with local magic 
    operations (blue gates).}
\end{figure}

We start from the observation that any universal quantum circuit, like the example shown in Fig. \ref{fig:stab_ent}, can be 
visualized as alternating Clifford unitary blocks and single-qubit rotations 
around Pauli axes:
\begin{equation}
U = C_0 R_1 C_1 R_2 C_2 \cdots R_\numrotdecomp C_\numrotdecomp,
\label{eq:clifford_magic_decomp}
\end{equation}
where each $C_j$ is a potentially \textit{non-local} Clifford unitary, each magic gate 
$R_j = \exp(-i\theta_j \sigma_{\mu_j} / 2)$ is a rotation by angle $\theta_j$ 
around axis $\mu_j \in \{X, Y, Z\}$ on a single qubit (or a \textit{local} 
rotation) and $N_{\rm nC}$ is the number of non-Clifford gates appearing in the circuit. 

A second central observation is that conjugation by a non-local Clifford $C$ 
will in general map a local Pauli $P$ to a non-local Pauli $C P C^\dagger$. For example, 
consider a local rotation on qubit 2 of a 5-qubit system, generated by the tensor product
\begin{equation}
     I \otimes I \otimes Z \otimes I \otimes I \equiv IIZII,
\end{equation}
so that the rotation acts as
\begin{equation}
    R = \exp\!\left[-i\frac{\theta}{2}\, IIZII \right].
\end{equation}
Applying a $\mathrm{CNOT}_{02}$ (control on qubit 0, target on qubit 2) 
followed by a Hadamard on qubits 0 and 2, the generator transforms under 
conjugation as
\begin{equation}
    IIZII \xrightarrow{\mathrm{CNOT}_{02}}
    ZIZII
    \;\xrightarrow{H_0 \otimes H_2} XIXII,
\end{equation}
where $Z$ on the target backpropagates to the control qubit via 
$Z_{\rm t} \to Z_{\rm c} Z_{\rm t}$, and $H Z H^\dagger = X$ maps each 
remaining $Z$ to $X$. The originally local rotation $\exp(-i\theta\, Z_2/2)$ 
is thus conjugated into the genuinely non-local two-body interaction 
$\exp\!\left(-i\theta\, X_0 \otimes X_2 / 2\right)$, which generates 
entanglement between qubits 0 and 2.

This motivates moving Clifford gates past the rotation operators via 
conjugation:
\begin{equation}
C_j R_j C_j^\dagger = \tilde{R}_j,
\label{eq:clifford_conjugation}
\end{equation}
where $\tilde{R}_j = \exp(-i\theta_j P_j / 2)$ and 
$P_j = C_j \sigma_{\mu_j} C_j^\dagger$ is now a (possibly non-local) Pauli 
string. By repeatedly applying this conjugation, any Clifford block can be 
``pushed through'' the rotation to its right, absorbing the Pauli-string 
rotation without changing the unitary:
\begin{equation}
U = ( C_1 R_1 C_1^\dagger ) ( C_1 C_2 R_2 C_2^\dagger C_1^\dagger)\cdots
  = \tilde{R}_1\tilde{R}_2\cdots\tilde{C},
\label{eq:clifford_decomposition}
\end{equation}
where $\tilde{R}_j$ are the dressed rotations and $\tilde{C}$ is the 
total residual Clifford evolution.

At this point it is worth pausing to understand \textit{why} this rewriting is advantageous for classical simulation. 

Indeed Eq.~\eqref{eq:clifford_decomposition} cleanly separates two distinct roles that the Clifford layers are responsible for: \textit{i)} spreading entanglement through the state \emph{after} any magic rotation is applied, and \textit{ii)} dressing the Pauli axes along which the magic rotations act on. 
The stabilizer MPO formalism takes advantage of this separation, adressing each point with the most suitable simulation technique. 

In particular, role \textit{i)} is handled by a stabilizer simulator, which tracks Clifford evolution in polynomial time in Heisenberg picture (see Sect. \ref{sec:mixed_pict}), while role \textit{ii)} is handled analytically and exactly, by replacing the Pauli generator of each $\tilde{R}_j$. What remains to simulate is only the sequence  $\tilde{R}_1 \tilde{R}_2 \cdots$, each of which, as we show in the next section, \emph{always} admits a bond-dimension-2 MPO representation, and can therefore efficiently handled by a tensor network approach. 
In this way, the entanglement injected into the state by the full sequence $\tilde{R}_1 \cdots \tilde{R}_\numrotdecomp$, lacking most entangling gates, grows slower than conventional techniques, allowing the MPS simulation to remain accurate at fixed bond dimension $\chi$ for substantially longer times compared to direct application of $U$. This is well explained and documented in the original work of Mello et al.~\cite{Mello_2024}, where some dynamics examples are also showcased.

\subsection{Tensor network representation}
\label{sec:hsmpo_tn}
The dressed operators $\tilde{R}_j$ admit compact Matrix Product Operator 
(MPO) representations with bond dimension exactly two. Specifically, any 
Pauli-string rotation
\begin{equation}
\exp\left[-i\frac{\theta}{2} \Sigma_\gamma\right] 
  = \cos(\theta/2)\, I - i \sin(\theta/2) \Sigma_\gamma,
\label{eq:pauli_rotation_mpo}
\end{equation}
where $\Sigma_\gamma = \sigma^{\gamma_1} \otimes \sigma^{\gamma_2} \otimes 
\cdots \otimes \sigma^{\gamma_\numqubits}$ is a tensor product of single-site Pauli 
operators, can be written as a bond-2 MPO. To see this explicitly, note that 
Eq.~\eqref{eq:pauli_rotation_mpo} is a linear combination of exactly two 
operators: the identity $I^{\otimes \numqubits}$, which is itself a product operator 
with bond dimension one, and the Pauli string $\Sigma_\gamma$, which is also 
a product operator with bond dimension one. Their linear combination can 
therefore be written as a single MPO with bond dimension two, whose local 
tensors at each site $\qubitidx$ take the block-diagonal form
\begin{equation}
W_\qubitidx = \begin{pmatrix} \phi_0 \, \sigma^0 & 0 \\ 0 & \phi_1 \, \sigma^{\gamma_\qubitidx} \end{pmatrix},
\label{eq:mpo_local_tensor}
\end{equation}
where $\phi_0 = \cos^{1/\numqubits}(\theta/2)$ and $\phi_1 = (-i\sin(\theta/2))^{1/\numqubits}$ 
are the site-local factors of the two coefficients, distributed evenly across 
all $\numqubits$ sites, and $\sigma^0 \equiv I$ is the single-site identity. The full 
operator is recovered by contracting the auxiliary indices across all sites:
\begin{equation}
\tilde{R}_j = W_1 W_2 \cdots W_\numqubits,
\end{equation}
where the matrix multiplication acts on the bond (auxiliary) indices, while 
the Pauli matrices act on the physical indices. The key observation is that 
this bond dimension of two is an \textit{exact} representation, and it is 
entirely independent of how many qubits the string $\Sigma_\gamma$ acts on 
non-trivially. A weight-$\numqubits$ Pauli string rotation, spanning the entire 
system, costs no more to represent as an MPO than a single-qubit rotation. 
It is precisely this property that makes the stabilizer MPO decomposition 
efficient: regardless of how non-local the dressed Pauli string $P_j$ becomes 
after Clifford conjugation, each $\tilde{R}_j$ retains a bond-2 MPO structure, 
and the total bond dimension of the evolved state grows only through the 
sequential application of these rank-2 layers, at a rate controlled 
by the angles $\{\theta_j\}$.

\subsection{Mixed picture evolution of HSMPO}\label{sec:mixed_pict}

In this context, we are interested in computing expectation values of linear combinations of Pauli observables like
\begin{equation}
O = \sum_i c_i P_i,
\end{equation}
where $P_i$ and $c_i$ are a Pauli strings involving any number of qubits of our system and $c_i$ are their coefficients respectively. If we consider the circuit decomposition introduced in Eq.~\eqref{eq:clifford_decomposition}, the expectation value of $O$ over the state prepared by $U$ can be computed as:
\begin{equation}
\begin{split}
&\langle O \rangle = \langle \psi | U^\dagger O U |\psi\rangle = \\
&= \langle \psi | \tilde{R}_1^\dagger \dots \tilde{R}_\numrotdecomp^\dagger \tilde{C}^\dagger O \tilde{C} \tilde{R}_1 \dots \tilde{R}_\numrotdecomp |\psi\rangle.
\label{eq:expectation}
\end{split}
\end{equation}
HSMPO adopts a mixed \textit{forward-backward} picture for the time evolution implementation, motivated by the decomposition introduced in Eq.~\eqref{eq:clifford_decomposition}.
On one hand, Clifford operations are applied in the Heisenberg picture, acting on observables. This is why we often refer to
these parts of the evolution as \textit{backpropagations}; an initial observable $O$ is evolved into a new observable $O'$ through the action of $\numcliff$ Clifford operations via tableau arithmetic, according to the stabilizer formalism. This operation has a computational cost which scales as $O(\numcliff\,\numqubits)$, where $\numqubits$ is the number of qubits \cite{Aaronson_2004, Gidney2021stimfaststabilizer}.

On the other hand, non-Clifford operations are implemented in Schrödinger picture, namely acting on the state with a \textit{forward} evolution. In this non-Clifford evolution, each gate $\tilde{R}_j$ is represented as an MPO, as given in Eq.~\eqref{eq:pauli_rotation_mpo}, and contracted directly with the MPS encoding the initial state. After each contraction, the bond dimension of the resulting state is truncated to remain below a prescribed maximum value $\chi$. Starting from an initial state $|\psi\rangle$, the gates $\tilde{R}_j$ are applied iteratively, with a compression step following each contraction. This approach leads to a  computational complexity which scales with the maximum bond dimension and the number of magic operations $\nummagic$ as $O(\chi^3 \nummagic)$. After the evolution steps, the Pauli string corresponing to the evolved operator $O'$ is applied to the evolved MPS $|\psi\rangle$, and the expectation value is computed as a tensor network contraction:
\begin{equation}
\begin{aligned}
\langle O \rangle 
&= \langle \psi | (\tilde{R}_1^\dagger \cdots \tilde{R}_\numrotdecomp^\dagger)(\tilde{C}^\dagger O \tilde{C})(\tilde{R}_1 \cdots \tilde{R}_\numrotdecomp) |\psi\rangle \\
&= \langle \psi' | U^\dagger O' U |\psi' \rangle.
\end{aligned}
\label{eq:contraction}
\end{equation}

It is worth noting that the two components of this scheme carry different levels of exactness. The stabilizer evolution is exact, introducing no approximation error. The tensor network evolution, by contrast, is approximate: truncating the bond dimension limits the maximum entanglement the simulation can faithfully represent. This hybrid strategy is not unique to the present work; related approaches combining exact Clifford treatment with approximate tensor network methods have been explored in the literature \cite{Begusic2024}.

\section{The \texttt{mpstab} simulator}
\label{sec:mpstab}

To turn the HSMPO framework into a practical tool, we developed \texttt{mpstab}, 
an open-source Python package that implements hybrid stabilizer-tensor network simulation. Find the source here: \\

\href{https://github.com/MatteoRobbiati/mpstab}{\faGithub\,\, https://github.com/MatteoRobbiati/mpstab.} \\

\begin{figure}[ht]
    \centering
    \includegraphics[width=1\linewidth]{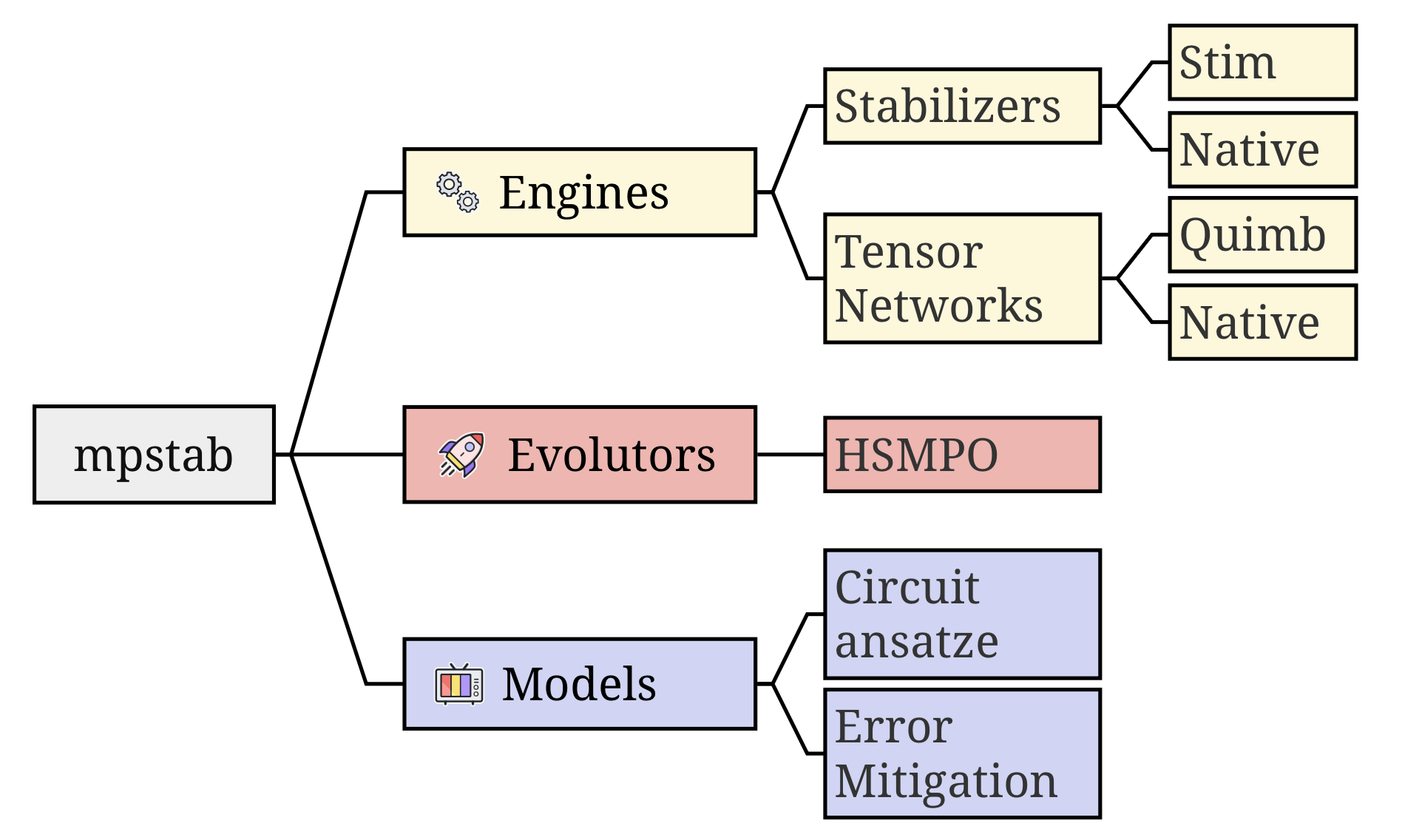} 
\caption{\label{fig:mpstab_scheme} Schematic representation of \texttt{mpstab 0.1.0}.}
\end{figure}

The library is designed to integrate seamlessly with Qibo~\cite{Qibo}, 
a high-level quantum computing framework, allowing users to express quantum circuits 
in familiar syntax while leveraging the HSMPO method to simulate more complicated circuits.

As summarized in Fig.~\ref{fig:mpstab_scheme}, the package is structured into three main modules: \textit{i)} the engines, deputed to the execution of the main operations described in the previous sections, \textit{ii)} the evolutors, namely our main interface, where the HSMPO formalism takes the form of a class and can be used to perform simulations. Finally, \textit{iii)} the models module contains a few built-it quantum circuit ansatze and some higher-level functionalities, such as experimental error mitigation methods relying on the HSMPO formalism or some relevant metrics, such as exact or stochastic estimate of the Stabilizer Rényi Entropy~\cite{Haug_2023}. These components represent the back-bone structure of \mpstab\, and will be expanded in future releases including new features, such as noise simulation. \\

At the user-facing level sits the \texttt{HSMPO} class, which orchestrates the entire 
workflow: it partitions an input circuit into Clifford and non-Clifford blocks, 
conjugates magic rotations with preceding Clifford gates to ensure Pauli structure 
is preserved, manages the MPS state evolution, applies truncation, and finally 
evaluates expectation values with automatic fidelity tracking. This abstraction 
insulates users from low-level technical details while maintaining the flexibility 
to swap engines and control key parameters such as the maximum bond dimension. \\

We provide our own tensor network and stabilizer simulators implementations, together with a Quimb~\cite{quimb} engine (for TN simulations) and a Stim~\cite{stim} one (for Stabilizers).

\subsection{Basic usage and API}
As anticipated, our main focus is providing a simple framework to simulate a quantum circuit in HSMPO mode. For this reason, we delegate the quantum circuit and observables syntaxes to Qibo~\cite{Qibo}.

After defining a quantum circuit using one of our built-in models or using the Qibo interface, one instantiates an 
\texttt{HSMPO} object as shown in the following code snippet
\vspace{0.3cm}
\begin{lstlisting}
from qibo import Circuit, gates
from mpstab.evolutors.hsmpo import HSMPO
from mpstab.models.ansatze import HardwareEfficient

# Create a parameterized circuit using a built in model
circuit = HardwareEfficient(nqubits=5, nlayers=3)

# or the Qibo syntax
circuit = Circuit(5)
for q in range(5):
    circuit.add(gates.RY(q=q, theta=0.34))
circuit.add(CZ(0, 3))
circuit.add(CZ(4, 1))

# Initialize simulator chosing the bond dimension
sim = HSMPO(circuit, max_bond_dimension=64)
\end{lstlisting}
\vspace{0.3cm}
When instantiating an \texttt{HSMPO} object, one can set the maximum bond dimension of the MPS representation. 


The evolution of the MPS through the series of the dressed rotations $\tilde{R}$ is computed once, and then cached to speed up the cost of computing multiple expectation values, as it happens when we consider an observable composed of many terms. An example of expectation value calculation is shown in the following code snippet.
\vspace{0.3cm}
\begin{lstlisting}
from qibo.symbols import X
from qibo.hamiltonians import SymbolicHamiltonian 

# Define an arbitrary symbolic hamiltonian
ham_form = 0
for q in range(5):
    ham_form += X(q%5) * X((q+1)%5) 
hamiltonian = SymbolicHamiltonian(ham_form)

# Compute the expectation value using HSMPO
sim.expectation(observable=hamiltonian)

# Expectation of the same observable using less magic circuit than the original 
sim.expectation_from_partition(
    observable=hamiltonian,
    replacement_probability=0.7,
    replacement_method="random",
)
\end{lstlisting} 
\vspace{0.3cm}
In the last block of code, we show a particular feature of \texttt{mpstab}, namely the expectation value computed over a processed version of the original circuit. If this method is called, \texttt{HSMPO} takes the original quantum circuit and replaces its magic gates (local rotations around general angles) with Clifford operations with an arbitrary chosen probability. This is done by setting the rotation angle to be a multiple of $\pi/2$. This feature is particularly useful when implementing data-driven error mitigation techniques or when exploring the simulator performances in various regimes of magic-to-stabilizerness ratio.

\subsection{Flexibility and modularity}
A key design principle of \texttt{mpstab} is modularity. Users can select different 
stabilizer and tensor network engines independently. For instance, one might use 
\texttt{StimEngine} for fast Pauli propagation paired with \texttt{QuimbEngine} 
configured on its Jax backend to enable automatic differentiation in case, for example, an HSMPO model is requested within a variational optimization pipeline. We propose this flexibility in order to consent the implementation of dedicated engines, which can be effortlessly integrated in our workflow. One can thus choose among the available engines and set them as shown in the following code snippet. 

\begin{lstlisting}
from mpstab.engines import StimEngine, QuimbEngine

# Choose engines independently
sim.set_engines(
    stab_engine=StimEngine(),
    tn_engine=QuimbEngine(backend="jax")
)
\end{lstlisting}

So far, the modular structure of \mpstab\ has been following a specific HSMPO structure, having stabilizer engines executing the backpropagation of Pauli strings and tensor network engines implementing the tensor network representation of the remaining part of the evolution. 

However, although this was the first mechanism implemented, we see significant room for exploration in terms of the engines used. For example, one could derive the analytical form of non-local rotations using a stabilizer engine and then attempt to synthesize a circuit in the style of digital quantum computing starting from there. Alternatively, one could consider using more complex tensor network ansatze or alternative hybrid approaches such as Pauli Propagation~\cite{Rudolph_2025}. A similar solution has indeed been implemented in \cite{pauli-prop}.

\subsection{Integration with existing workflows}
To enable a further modular usage of \texttt{mpstab}, we enable it to be registered as a Qibo backend. This can be easily done by using the dedicated backends API implemented in Qibo. We provide a simple working example in our documentation at~\cite{mpstab_qibo}.

The ability to use the HSMPO formalism whilst leveraging all the tools of a rich and well-established framework such as Qibo significantly speeds up the exploration of new solutions, such as the use of HSMPO in variational contexts like those for which QiboML was introduced~\cite{qiboml}.

\begin{figure*}[ht]
    \centering
    \includegraphics[width=0.85\linewidth]{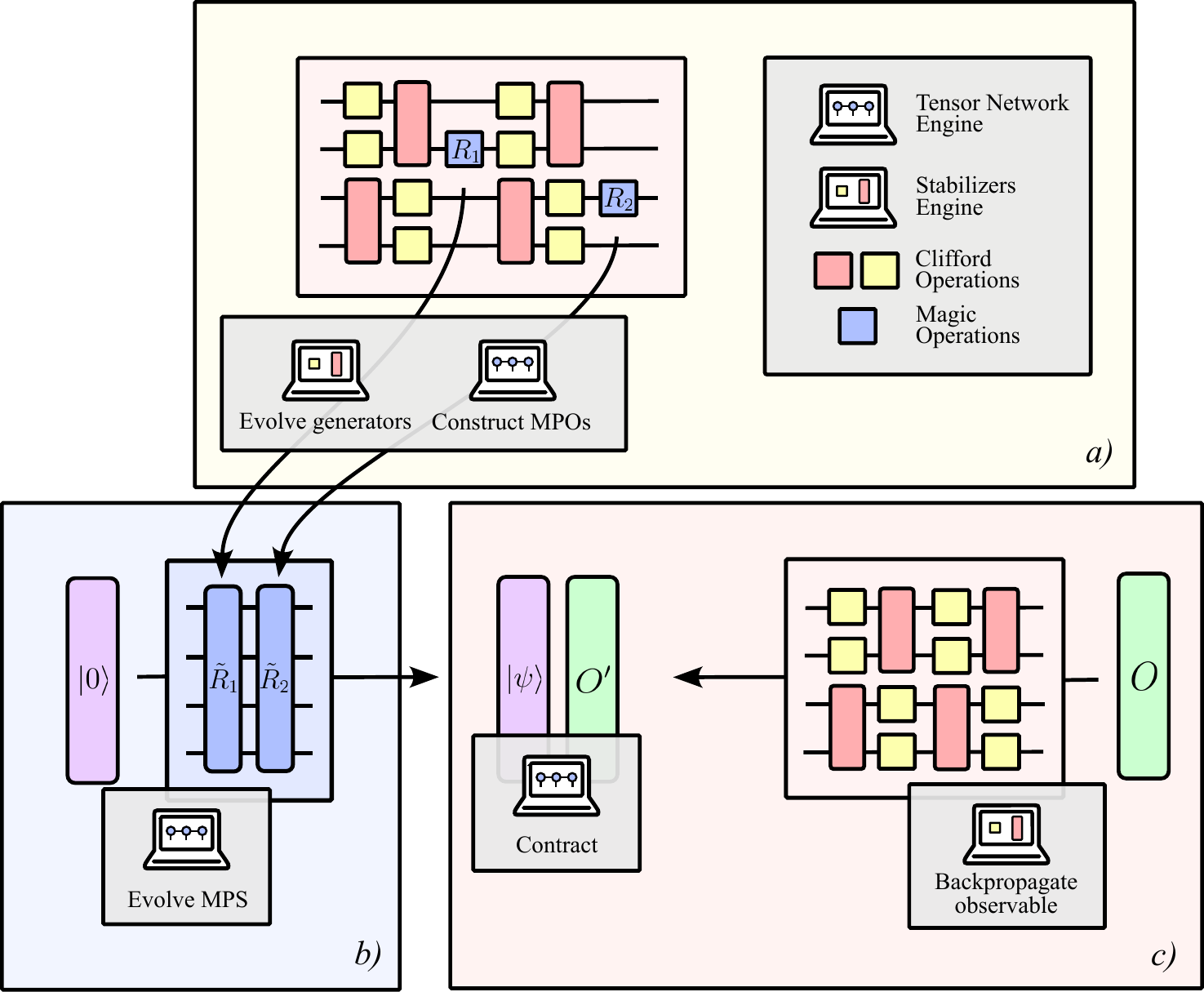} 
\caption{\label{fig:mpstab_pipeline} \texttt{mpstab}'s operational workflow when computing the expectation value of an observable $O$ over the state represented by an HSMPO. A tensor network and a stabilizers engines cooperate to prepare the evolved MPS and use it to perform the contraction with the given observable.}
\end{figure*}

More remarkably, Qibo offers a simple interface to real quantum devices~\cite{qibolab}, and its backends can be easily constructed and used in synergy to propose hybrid quantum-classical pipelines. \\

Modularity plays a crucial role in nowadays quantum computing applications, where quantum and classical components act as hardware accelerators within hybrid contexts. 

\subsection{Computational complexity}
\label{sec:complexity}

The computational cost of computing an expectation value with \mpstab\ is 
determined by the interplay between tensor network and stabilizer operational 
engines. To give a graphical intuition of the computational workflow, we 
support the next discussion with the schematic representation shown in 
Fig.~\ref{fig:mpstab_pipeline}.

The whole computational workflow can be divided into three steps: 

\begin{itemize}
    \item[\textit{a)}] \textbf{Operation setup}: identification of the Clifford 
    blocks and magic rotations, and construction of the dressed rotations.
    \item[\textit{b)}] \textbf{MPS evolution}: an initial state, represented as 
    an MPS, is evolved through the layer of dressed rotations.
    \item[\textit{c)}] \textbf{Expectation value evaluation}: the target 
    observable is evolved under the cumulative Clifford layer and contracted 
    with the MPS obtained at step \textit{b}.
\end{itemize}

Let $\numqubits$ be the number of qubits and $\numgates$ be the total number 
of gates, where $\numgates = \numcliff + \nummagic$ represents the sum of 
Clifford and non-Clifford operations. \\

\paragraph{Operation setup.}
This first operation consists in locating the position of magic gates in the 
queue of operations composing the quantum circuit, so that we can identify 
the Clifford blocks and the local rotations mentioned in 
Eq.~\eqref{eq:clifford_magic_decomp}. This requires a single pass over the 
gate list, scaling as $O(\numgates)$. For every magic gate $R_i$ in the 
decomposition of Eq.~\eqref{eq:clifford_magic_decomp}, we must evolve the 
rotation axis through the non-local Clifford circuit built from all Clifford 
gates preceding the target local rotation $R_i$ in the original circuit. In 
the symplectic representation of the stabilizer formalism, each elementary 
1- or 2-qubit Clifford gate updates only $O(1)$ entries of the Pauli string 
encoding the rotation axis. Consequently, the computational cost of preparing 
the generator of the dressed rotation $\tilde{R}_i$ scales as $O(N_{C,i})$, 
where $N_{C,i}$ is the number of Clifford gates preceding the $i$-th rotation.

Considering the extreme (worst) case where each rotation is dressed under the action 
of all the Clifford gates populating the circuit, the computational cost of 
this step scales as $O(\nummagic\,\numcliff)$. \\

\begin{figure*}[ht]
    \centering
    \includegraphics[width=1\linewidth]{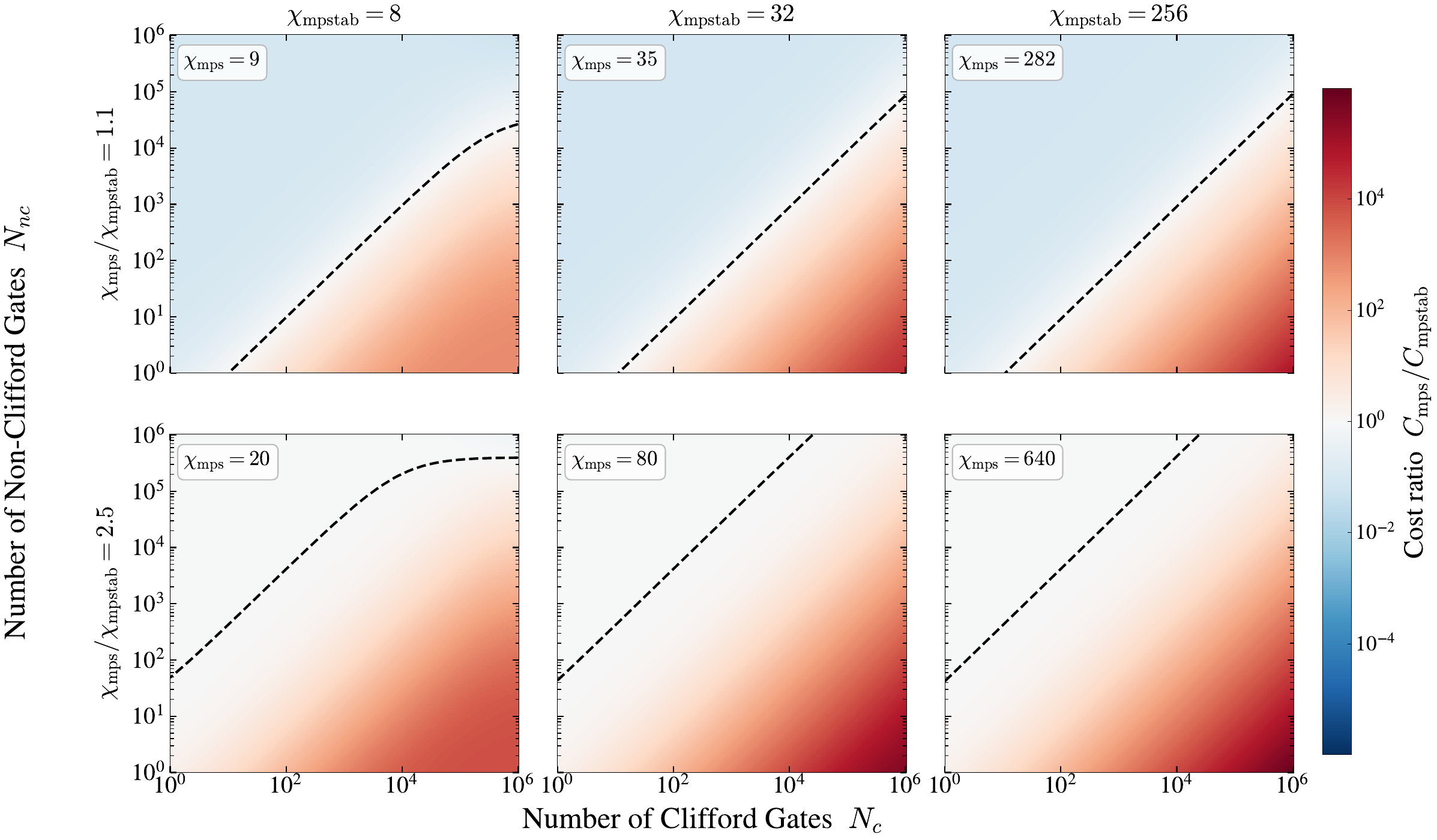} 
    \caption{\label{fig:cost_advantage_grid}Cost ratio 
$C_{\rm mps}/C_{\rm mpstab}$ as a function of Clifford ($\numcliff$) and 
non-Clifford ($\nummagic$) gate counts. Red regions indicate where \mpstab\ 
is cheaper than standard MPS; blue regions indicate the opposite. The dashed 
curve marks the iso-cost boundary $C_{\rm mps} = C_{\rm mpstab}$. Columns 
fix the \mpstab\ bond dimension 
$\chi_{\rm mpstab} \in \{8, 32, 256\}$, spanning strongly- to 
weakly-compressed hybrid representations; rows fix the ratio 
$\chi_{\rm mps}/\chi_{\rm mpstab} \in \{1.1, 2.5\}$ between the two methods, 
corresponding respectively to a weak and a strong fidelity advantage of the 
hybrid method. The leftmost column exhibits the arched regime, where the 
stabilizer overhead $\nummagic\,\numcliff$ saturates the boundary at high 
$\numcliff$; the middle and right columns exhibit the diagonal regime, 
where the boundary is set entirely by the cubic bond-dimension gap. System 
size is $\numqubits = 50$ with worst-case interaction range 
$\maxintlen = \numqubits$.}
\end{figure*}

\paragraph{MPS evolution.}
Once the generators of the dressed rotations are constructed, these non-local 
operations are applied as MPOs to the initial state, which is written in the 
form of an MPS. Contracting an MPO tensor (with bond dimension $D=2$, as 
shown in Sec.~\ref{sec:hsmpo_tn}) with an MPS tensor of bond dimension 
$\chi$ scales as $O(d^2 D^2 \numqubits \chi^2) = O(16\,\numqubits \chi^2)$ 
for physical dimension $d=2$. Following the contraction, a singular value decomposition (SVD)-based 
compression is performed to keep the bond dimension bounded by $\chi$; the 
SVD of each local matrix of size $(d\chi D)\times(\chi D)$ costs 
$O(d D^3 \chi^3)$, giving a total 
$O(d D^3 \numqubits \chi^3) = O(16\,\numqubits \chi^3)$. The full evolution 
of a magic block is therefore dominated by the SVD step and scales as 
$O(16\,\numqubits \chi^3)$. 

Referring back to the two sub-routines mentioned at the beginning of this 
section, the MPS evolution is computed in \mpstab\ through steps \textit{a} 
and \textit{b}, while the expectation value calculation corresponds to step 
\textit{c}. \\

\paragraph{Expectation value of a target observable.}
The evaluation of the expectation value of a target observable $O$ requires 
backpropagating the observable through the full Clifford circuit, obtained by 
concatenating all the Clifford blocks identified at step \textit{a}. By the 
same argument used for the dressing, this operation costs $O(\numcliff)$ per 
Pauli term in the observable; assuming an observable with $O(\numqubits)$ 
Pauli terms (as is typical for $k$-local Hamiltonians) yields a total cost 
of $O(\numqubits\,\numcliff)$. The resulting evolved Pauli string, 
represented as a product operator, is then contracted with the evolved MPS 
$|\psi'\rangle$. This final contraction between the Pauli string and the 
state scales as $O(d^2 \numqubits \chi^2) = O(4\,\numqubits \chi^2)$, 
ensuring that the measurement phase remains computationally efficient. \\

\subsubsection{Comparison with standard MPS}
\label{sec:comparison}

In standard matrix product state (MPS) simulation, gates are applied 
sequentially to the state representation. As discussed 
in~\cite{jozsa2006simulationquantumcircuits}, single-qubit gates can be 
applied to an MPS state with cost $O(\chi^2)$, adjacent two-qubit gates with 
cost $O(\chi^3)$, and non-adjacent two-qubit gates acting across spatial 
distance $r$ require swap sequences costing $O(r\chi^3)$. For a circuit of 
$\numgates$ gates with maximum interaction range $\maxintlen$, the total 
upper-bound complexity is $O(\numgates\,\maxintlen\,\chi^3)$, with a final 
contraction of the observable with the state adding 
$O(d^2\numqubits\chi^2) = O(4\numqubits\chi^2)$ for physical dimension $d=2$.

Letting $\numgates = \numcliff + \nummagic$ denote the total gate count, 
split into $\numcliff$ Clifford and $\nummagic$ magic gates, and assuming 
worst-case non-locality $\maxintlen = \numqubits$, the standard MPS cost 
reads:
\begin{equation}
    C_{\rm mps} = (\numcliff + \nummagic) \, \numqubits \, \chi_{\rm mps}^3 
                + 4 \, \numqubits \, \chi_{\rm mps}^2,
\end{equation}
where the \textit{mps} label on the bond dimension emphasises that MPS and 
HSMPO representations may require different bond dimensions to achieve the 
same accuracy in the comparison with \mpstab.

Aggregating the costs derived in Sec.~\ref{sec:complexity}, setup, 
stabilizer backpropagation of the observable, dressing of the magic 
rotations, MPS evolution through the dressed rotations, and final 
measurement, the \mpstab\ cost is:
\begin{equation}
\begin{split}
C_{\rm mpstab} = {} & \numgates + \numqubits \numcliff + \nummagic \, \numcliff \\
                    & + 16 \, \nummagic \, \numqubits \, \chi_{\rm mpstab}^3 
                    + 4 \, \numqubits \, \chi_{\rm mpstab}^2,
\end{split}
\end{equation}
where the $\numqubits \numcliff$ term accounts for the backpropagation of an 
observable with $O(\numqubits)$ Pauli terms through all Clifford operations, 
and $\nummagic \, \numcliff$ collects the cost of dressing each of the 
$\nummagic$ magic rotations under all preceding Clifford gates.

As a first performance metric, we analyse the ratio $C_{\rm mps}/C_{\rm mpstab}$ as a function of the non-Clifford and Clifford gates of the circuit. Both $\numcliff$ and $\nummagic$ are varied from zero to one million of gates. 

Fig.~\ref{fig:cost_advantage_grid} maps this ratio across the 
circuit-composition space for six pairs of maximum bond dimensions 
$(\chi_{\rm mps}, \chi_{\rm mpstab})$. The three values 
$\chi_{\rm mpstab} \in \{8, 32, 256\}$ span from strongly compressed to 
weakly compressed hybrid representations, while the two ratios 
$\chi_{\rm mps}/\chi_{\rm mpstab} \in \{1.1, 2.5\}$ encode the 
bond-dimension saving that \mpstab\ achieves at fixed fidelity; a larger ratio corresponds to a 
stronger fidelity advantage of the hybrid method.

Two regimes are visible. For moderate $\chi_{\rm mpstab}$ (left column), the iso-cost contour has an arched shape: \mpstab\ is 
cheaper in a finite window of $\numcliff$ that closes at very large Clifford 
counts, where the $\nummagic \, \numcliff$ contribution to the \mpstab\ cost 
eventually dominates. For larger $\chi_{\rm mpstab}$ (middle and right column) this term 
is subleading, and the boundary becomes a straight diagonal in log--log 
scale, set primarily by the cubic bond-dimension gap 
$(\chi_{\rm mps}/\chi_{\rm mpstab})^3$.

Increasing the ratio $\chi_{\rm mps}/\chi_{\rm mpstab}$ (top to bottom row) 
systematically expands the \mpstab-favourable region, reflecting the cubic 
scaling with bond dimension of the standard method. Overall, the \mpstab\ 
advantage is most pronounced at low $\nummagic$ and high $\numcliff$, where 
the stabilizer formalism absorbs the bulk of the entanglement structure and 
the non-Clifford overhead remains modest.

\section{Benchmarks}

As widely discussed, HSMPO is introduced as a tool to simulate longer dynamics 
and more entangled operations. While this is true on a theoretical perspective, 
implementing this in practice introduces a series of overheads and complexities 
that makes non-trivial the task of identifying when our simulator is supposed 
to be the to-go choice among the rich pletora of alternatives. 

With this in mind, and considering the task of computing expectation values, 
we want to answer here the question: \textit{for which kind of circuits 
\texttt{mpstab} is the best available simulator?} 

It is natural for us to structure these benchmarks \textit{against} a pure 
tensor network simulator. This allows the analysis of \texttt{mpstab}'s 
performances to be the fairest as possible considering two aspects: 
\textit{i)} the alternative method (TN) doesn't present any overhead and 
\textit{ii)} using Quimb, we can easily force the pure tensor network 
simulation and the \texttt{mpstab}'s tensor network engine to follow the same 
configuration: contractions optimizer, backend, etc.

\subsection{Benchmark setup}
\label{sec:bench_setup}

The two main characteristics impacting on the analysis of the performance are 
the amount of \textit{entanglement} and \textit{magic} in the circuit. We 
therefore design an \textit{ad hoc} ansatz that provides precise, independent 
control over these two resources, and complement it with a metrology protocol 
that certifies the accuracy of \mpstab\ without ever needing access to the 
exact state. \\

\begin{figure}[ht]
    \centering
    \includegraphics[width=1\linewidth]{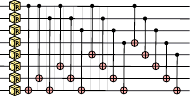} 
    \caption{
    \label{fig:random_circ}    
    One layer of the quantum circuit used for benchmarking. First we have a layer of $\numqubits$ single qubit rotations, a fraction $1-\beta$ of which, will be Clifford, the rest will be random. The interaction part consists of a CNOT ladder containing all possible CNOT gates from interaction length $\numqubits-1$ down to $k=5$.
    }
\end{figure}

\paragraph{Ansatz.}

Each layer of the benchmark ansatz, illustrated in Fig.~\ref{fig:random_circ}, 
consists of $\numqubits$ single-qubit rotations followed by a CNOT ladder of 
tunable depth. The rotations are drawn so that a fraction $\mrr$ of them is 
non-Clifford, with random rotation angles, and the remaining $1-\mrr$ is 
Clifford, with angles multiple of $\pi/2$, so that the magic content of the 
layer can be dialed continuously. The interaction block is a ladder of CNOT 
gates spanning interaction lengths from $\numqubits-1$ down to a cutoff $k$, 
yielding $k(k-1)/2$ entangling Clifford gates per layer; smaller $k$ means a 
deeper ladder and correspondingly more entanglement injected. Under this 
construction, the magic-to-Clifford ratio per layer reads
\begin{equation}
    \frac{\nummagic}{\numcliff} 
    = \frac{\numqubits\,\mrr}{\numqubits\,(1-\mrr) + k(k-1)/2},
    \label{eq:magic_cliff_ratio}
\end{equation}
so that the hyperparameters $(\mrr, k)$ span the two axes of the entanglement 
versus magic diagram of Fig.~\ref{fig:classical_simulators} in a controlled 
way. Throughout the benchmarks we tune the magic content using the 
quantity $\mrr$, which we dub \textit{magic retention}. \\

\paragraph{Certifying accuracy via a fidelity lower bound.}
Since we operate deep in the many-qubit regime, exact-state fidelities are 
inaccessible and we must rely on an intrinsic certificate of approximation 
quality. \mpstab\ tracks a fidelity lower bound throughout the simulation: 
at each MPS compression step, truncating the singular values below the bond 
dimension threshold discards a portion of the state norm. Since unitary 
evolution preserves the norm exactly, any deviation of the final MPS norm 
from unity is entirely attributable to truncation. The fidelity lower bound 
is therefore
\begin{equation}
    \mathcal{F}_{\rm lb} = \frac{\|\psi_{\rm trunc}\|^2}{\|\psi_0\|^2} 
    = \|\psi_{\rm trunc}\|^2,
    \label{eq:fidelity_lb}
\end{equation}
which provides a cheap and rigorous certificate that requires no access to 
the exact state. In the following, whenever we refer to ``fidelity'' we mean 
this lower bound; the same quantity is also tracked for the pure TN baseline 
for a fair comparison. \\

\paragraph{Protocol and parameter grid.}
The \mpstab\ package is focused on expectation value calculation. The inputs 
for a single test are the observable to be measured, the quantum circuit, 
the magic retention $\mrr$, and the maximum bond dimension $\chi_{\max}$. 
The outputs are the expectation value, the fidelity lower bound of 
Eq.~\eqref{eq:fidelity_lb}, and the execution time. For each configuration, 
both \mpstab\ and Quimb are executed under identical conditions: 
the pure Quimb tensor network simulation and the tensor network component 
of \mpstab\ share the same contraction algorithms and optimization settings, 
so that any observed difference in cost or accuracy is attributable to the 
hybrid formalism and not to implementation-level asymmetries. To collect 
statistics, each configuration is repeated $M = 20$ times with independent 
random initializations of the non-Clifford rotations. The hyperparameter 
values scanned in this study are collected in 
Table~\ref{tab:hyperparameters}. Results are reproducible using the 
\href{https://github.com/mattia-robbiano/mssim}{\texttt{mssim}} package; 
instructions are available in the 
\href{https://github.com/MatteoRobbiati/mpstab/tree/main/reproduce_results}{\mpstab\ 
repository}.

\begin{table}[h]
    \centering
    \begin{tabular}{ll}
    \hline \hline
    \textbf{Parameter} & \textbf{Values} \\
    \hline 
    Number of qubits $\numqubits$ & 20, 35, 50, 65, 80 \\
    Circuit depth $\numlayers$    & 1, 2, 3, 4, 5 \\
    Bond dimension $\chi_{\max}$  & 2, 4, 8, 16, 32 \\
    Magic retention $\mrr$      & 0, 0.2, 0.4, 0.6, 0.8, 1.0 \\
    \hline \hline
    \end{tabular}
    \caption{\label{tab:hyperparameters}Hyperparameter values scanned in the 
    benchmark simulations. Each configuration is repeated over $M=20$ random 
    initializations of the non-Clifford rotations.}
\end{table}

\subsection{Numerical results}
\label{sec:bench_results}

\begin{figure*}[ht]
    \centering
    \includegraphics[width=0.45\linewidth]{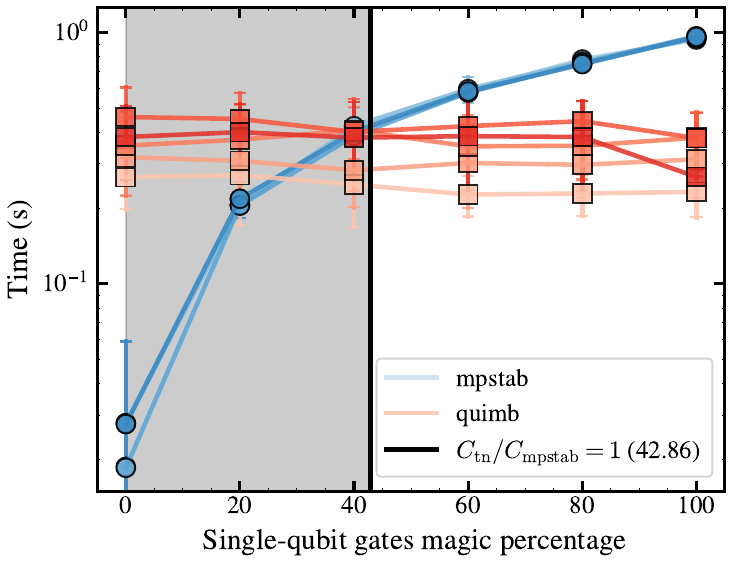} 
    \hfill
    \includegraphics[width=0.45\linewidth]{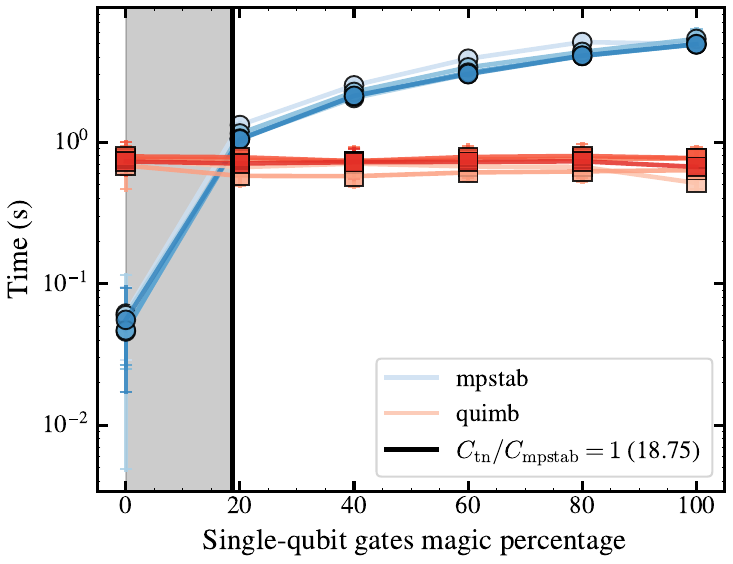} 
    \caption{\label{fig:execution_time_magic}Execution time as a function of 
    magic fraction with a single circuit layer and varying bond dimensions. 
    Blue and red lines correspond to \texttt{mpstab} and pure tensor network, 
    respectively. The vertical line marks the predicted crossover point 
    where computational costs are equivalent.
    \textit{Left:} $\numqubits = 35$ qubits.
    \textit{Right:} $\numqubits = 80$ qubits.}
\end{figure*}

\begin{figure*}[ht]
    \centering
    \includegraphics[width=0.45\linewidth]{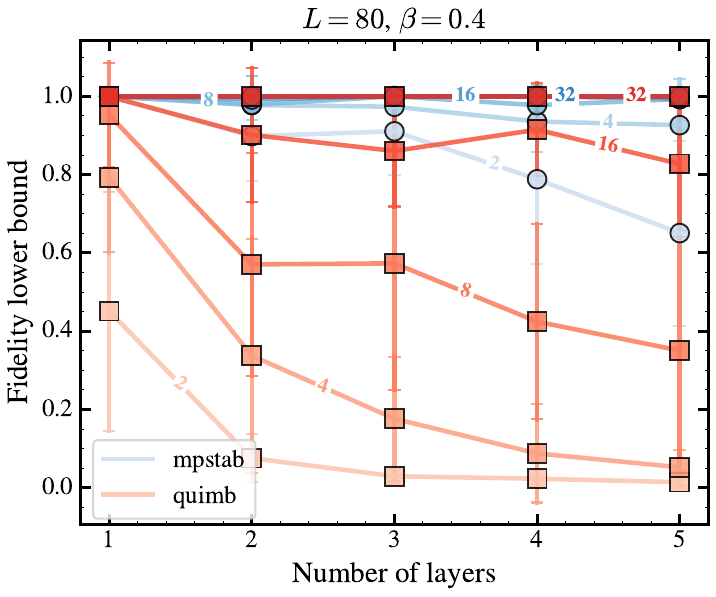} 
    \hfill
    \includegraphics[width=0.45\linewidth]{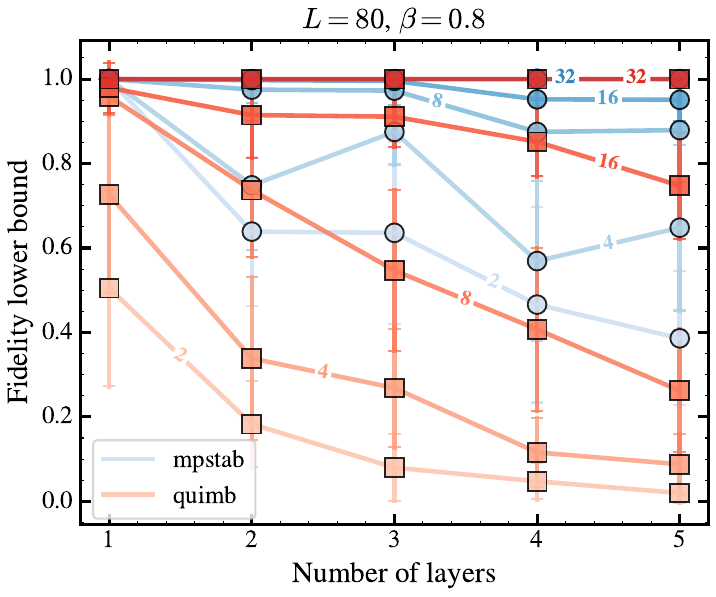} 
    \caption{\label{fig:fidelity_q80_scaling}Fidelity lower bound of 
    \texttt{mpstab} expectation value computations for 80-qubit circuits as 
    a function of circuit depth. Blue and red lines correspond to 
    \texttt{mpstab} and pure tensor network, respectively. Different shades 
    correspond to different values of $\chi_{\max}$, reported as labels on 
    the lines. Markers and error bars show median and MAD of fidelity values 
    across $M=20$ unitaries.
    \textit{Left:} $\mrr = 0.4$, moderate magic.
    \textit{Right:} $\mrr = 0.8$, high magic.}
\end{figure*}

\begin{figure*}[ht]
    \centering
    \includegraphics[width=0.45\linewidth]{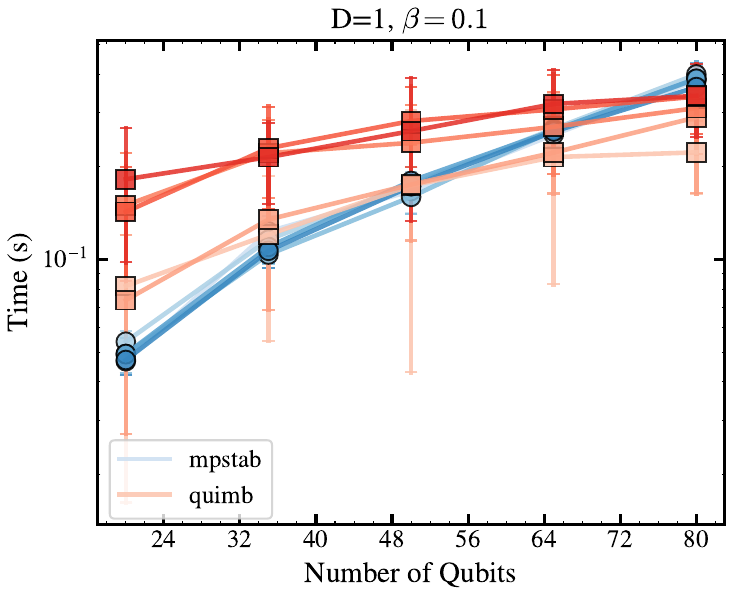} 
    \hfill
    \includegraphics[width=0.45\linewidth]{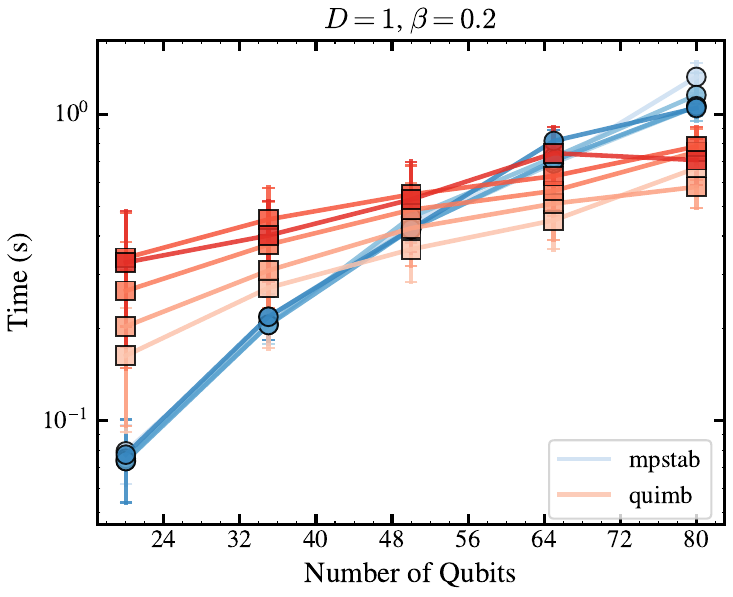} 
    \caption{\label{fig:execution_time_qubits}Execution time as a function of 
    number of qubits with a single circuit layer and varying bond dimensions. 
    Blue and red lines correspond to \texttt{mpstab} and pure tensor network, 
    respectively.
    \textit{Left:} $\mrr = 0.1$.
    \textit{Right:} $\mrr = 0.2$.}
\end{figure*}

We now walk through the three complementary views on \mpstab's performance 
provided by our benchmarks. The narrative follows the logic set up in 
Sec.~\ref{sec:complexity}: we first confront the analytical cost model 
against wall-clock measurements to validate it (Fig.~\ref{fig:execution_time_magic}), then dissect the two sources 
of \mpstab's advantage separately, namely its higher fidelity at fixed bond 
dimension (Fig.~\ref{fig:fidelity_q80_scaling}) and its wall-clock scaling 
with system size (Fig.~\ref{fig:execution_time_qubits}). \\

\paragraph{Validating the cost model.}
Fig.~\ref{fig:execution_time_magic} shows the wall-clock execution time of 
\mpstab\ and pure Quimb as a function of the single-qubit magic percentage, for two representative system sizes 
$\numqubits \in \{35, 80\}$ at fixed depth $\numlayers=1$. As the magic 
percentage increases from $0\%$ to $100\%$, \mpstab\ moves from a fully 
stabilizer-tractable regime, where it exploits polynomial-time Clifford 
propagation, to a regime dominated by dressed non-Clifford rotations that 
must all be applied as MPOs to the MPS. Correspondingly, its execution time 
grows monotonically with the magic fraction, while pure Quimb, which is 
insensitive to the Clifford versus non-Clifford distinction, shows an 
essentially flat cost across the whole range.

The two curves cross at a well-defined magic fraction, marked in the figure 
by a vertical line at the value predicted by setting 
$C_{\rm mps}/C_{\rm mpstab} = 1$ in the cost model of 
Sec.~\ref{sec:comparison}, Eqs.~(13) and (14). The predicted crossover falls 
inside the interval where the measured curves actually cross, both at 
$\numqubits=35$ and at $\numqubits=80$, confirming that the cost model of 
Sec.~\ref{sec:complexity} correctly predicts the order of magnitude of the 
crossover point, and therefore captures the leading-order scaling of the 
real implementation. This validation is important because it means the phase 
map of Fig.~\ref{fig:cost_advantage_grid}, drawn purely from the analytical 
model, can be trusted as a practical guide to when \mpstab\ is expected to 
outperform pure MPS. \\

\paragraph{Fidelity advantage at fixed bond dimension.}
Once the cost model is validated, it is worth asking \emph{why} \mpstab\ can 
be cheaper than a pure tensor network at all, given that the underlying 
tensor network engine is the same. The answer, anticipated in 
Sec.~\ref{sec:hsmpo}, is that Clifford conjugation absorbs a substantial 
portion of the entanglement structure of the circuit into the stabilizer 
formalism, so the MPS component of \mpstab\ has to represent only the 
entanglement generated by the dressed non-Clifford rotations. At fixed bond 
dimension, this translates directly into a higher accuracy.

Fig.~\ref{fig:fidelity_q80_scaling} makes this concrete for 80-qubit 
circuits, plotting the fidelity lower bound of Eq.~\eqref{eq:fidelity_lb} as 
a function of circuit depth, at multiple values of $\chi_{\max}$. At fixed 
$\chi_{\max}$, \mpstab\ (blue) is systematically above pure Quimb (red), and 
the gap widens with depth: as the circuit accumulates Clifford (thus entanglement), 
the pure TN representation degrades quickly while \mpstab\ absorbs it 
exactly in the stabilizer component. The advantage is most pronounced at 
moderate magic ($\mrr=0.4$, left panel) where the stabilizer component 
contains a substantial fraction of the dynamics, but persists even at high 
magic content ($\mrr=0.8$, right panel), where \mpstab\ still maintains 
significantly higher fidelity than the TN baseline at the same 
$\chi_{\max}$. This behavior is consistent with the picture underpinning 
the cost analysis of Sec.~\ref{sec:comparison}: reaching a target fidelity 
with \mpstab\ requires a strictly smaller bond dimension than with pure 
MPS, i.e., $\chi_{\rm mps} > \chi_{\rm mpstab}$, which is precisely the 
regime in which the phase map of Fig.~\ref{fig:cost_advantage_grid} 
favours the hybrid method. \\

\paragraph{Wall-clock cost and overhead across system sizes.}
The complementary question is how the two methods scale with system size in 
real wall-clock time. Fig.~\ref{fig:execution_time_qubits} shows the execution time as a function of the number of qubits, at two fixed magic retention values and single-layer depth. Both curves grow monotonically with system size $\numqubits$, with the $\mpstab$ wall-clock following a steeper path.
At low magic content ($\mrr=0.1$, left panel) \mpstab\ performs well for a larger number of qubits compared to case of high magic content ($\mrr=0.2$, right panel), where advantage is confined to smaller systems. This result highlights once again that the setting most favorable to $\mpstab$ that of low $\nummagic$. Indeed the number of magic gates $\nummagic \propto \mrr \numqubits$ is by construction proportional to the system size in this example, meaning that for fixed $\beta$, an increase in the system size also implies an increase in $\nummagic$, and thus the competitiveness of $\mpstab$ is reduced in accordance with the cost model of Sec.~\ref{sec:comparison}. Tuning the scaling of $\mrr$ with increasing $L$ is therefore crucial to maintain the advantage at scale.


\paragraph{Answer to the guiding question.}
Combining the three views, we can now return to the question that opened 
this section. \mpstab\ is the preferred simulator whenever \textit{(i)} the 
Clifford content is dominating
, so that the 
stabilizer formalism absorbs a substantial share of the entanglement and 
\textit{(ii)} the target fidelity requires a bond dimension high enough 
that the cubic-in-$\chi$ MPS cost dominates. This is precisely the region 
identified as \mpstab-favourable in the analytical phase map of 
Fig.~\ref{fig:cost_advantage_grid}, low $\nummagic$ and high $\numcliff$, 
and it corresponds, physically, to Clifford-dominated circuits with 
moderate but non-trivial non-Clifford injection. Outside this window, that 
is, very low Clifford content, very small systems, or bond dimensions small 
enough that the tensor-network cost is not yet the bottleneck, the pure TN 
baseline remains a better choice.


\section{Conclusions}

We have presented \mpstab, an open-source Python library implementing the 
hybrid stabilizer-MPO framework of Ref.~\cite{Mello_2024}. By treating 
Clifford gates exactly and encoding non-Clifford rotations as bond-2 MPO 
layers, HSMPO extends the reach of classical simulation beyond what 
tensor networks and stabilizer methods can achieve independently, and 
\mpstab\ turns this formalism into a practical tool.

Two design choices make \mpstab\ convenient for daily use. Its modular 
architecture cleanly separates the stabilizer and tensor network engines, 
so either component can be swapped transparently, for instance with Stim 
or Quimb. Integration with Qibo places \mpstab\ inside a broader ecosystem 
spanning circuit definition, variational optimization, and execution on 
real quantum hardware.

Our benchmarks locate the regime where the method shines: circuits with 
moderate magic and substantial Clifford content, where \mpstab\ reaches 
higher fidelity than pure tensor networks at the same bond dimension, 
translating into real savings once the comparison is drawn at matched 
accuracy. The analytical cost model of Sec.~\ref{sec:complexity} correctly 
predicts where the crossover happens, giving a concrete map of the 
parameter space in which \mpstab\ is the tool of choice.

Several directions remain open. On the algorithmic side, the Clifford 
blocks that precede a magic rotation shape the locality of the resulting 
dressed generator, and hence the bond-dimension growth of the MPS 
evolution; conversely, the Clifford operations sitting after all magic 
rotations dress the final observable, potentially turning a simple Pauli 
string into a complex sum of terms. A systematic study of how different 
Clifford substructures propagate axes and observables would help identify 
topologies that keep both objects as compact as possible. Recent work has 
begun to characterize this question directly, studying the disentangling 
power of Clifford transformations acting on tensor networks and proving 
fundamental limits on when a single qubit can be disentangled from an 
arbitrary non-Clifford rotation~\cite{Masot_Llima_2026}; extending this 
characterization to the dressed-rotation setting of HSMPO is a natural 
next step.

On the software side, the modularity of \mpstab\ invites new engines: 
Pauli propagation for the observable side, richer tensor network ansatze 
for the state side, or symbolic Clifford decomposers returning the 
dressed generators in analytical form. Each of these would probe a 
different corner of the entanglement versus magic diagram of 
Fig.~\ref{fig:classical_simulators}.

Finally, applications where circuits are structurally low in magic map 
naturally onto \mpstab. Learning-based error mitigation is a prime 
example: techniques like Clifford Data Regression train on surrogate 
circuits obtained by replacing most non-Clifford gates with Clifford 
ones, and are typically bottlenecked by the cost of generating those 
surrogates with exact simulators. \mpstab\ removes this bottleneck, 
allowing a controlled amount of magic to be retained while scaling to 
system sizes well beyond exact simulation, and its replacement 
functionalities are designed precisely for this use case. More broadly, 
we expect \mpstab\ to be a useful component in any hybrid 
quantum-classical pipeline where the classical side must simulate 
near-Clifford surrogates of quantum circuits. 

\section{Acknowledgments}
We thank Mario Collura, Stefano Carrazza, Cenk Tüysüz and Andrea Papaluca for useful discussions.
Mattia Robbiano and Giulio Crognaletti gratefully acknowledge the financial support of the Finnish Foundation for Technology Promotion. Giulio Crognaletti also acknowledges financial support from University of Trieste and INFN.
Michele Grossi is supported by CERN through the CERN Quantum Technology Initiative. Matteo Robbiati
was financially supported by the Knut and Alice
Wallenberg Foundation through the Wallenberg
Center for Quantum Technology (WACQT), the
Horizon Europe programme HORIZON-CL4-
2022-QUANTUM-01-SGA via the project 101113946 OpenSuperQPlus100 and the EuroHPC programme DIGITAL-EUROHPC-JU-2022-HPCQC-04-01-IBA via project 101159808 EUROQHPC-I.
\bibliographystyle{apsrev4-2}
\bibliography{references}

\end{document}